\RequirePackage[OT1]{fontenc}

\documentclass[conference]{IEEEtran}
\usepackage{cite}
\usepackage{amsmath,amssymb,amsfonts}
\usepackage{graphicx}
\usepackage{textcomp}
\usepackage[colorlinks=true,urlcolor=teal,
            linkcolor=teal,citecolor=teal]{hyperref}
\usepackage{lipsum}
\usepackage{xspace}
\usepackage{braket}
\usepackage{tikz}
\usepackage{mathtools}
\usepackage{comment}
\usepackage{placeins}
\usepackage{csquotes}
\usepackage{ifthen}

\newcommand{\eg}{\emph{e.g.}\xspace}
\newcommand{\ie}{\emph{i.e.}\xspace}
\newcommand{\etal}{\emph{et al.}\xspace}

\newcommand{\sgate}{\textsc{Swap} gate\xspace}
\newcommand{\sgates}{\textsc{Swap} gates\xspace}
\newcommand{\cd}{connectivity density\xspace}
\newcommand{\cds}{connectivity densities\xspace}

\newcommand{\CC}[1]{\ensuremath{\text{#1}}}

\newcommand{\suppweb}{\href{https://quantumdoubleblind.github.io/qsw_2023/}{supplementary website}\xspace}
\newcommand{\CXgate}{\ensuremath{\text{C\raisebox{0.08em}{--}}\!X}\xspace}

\newif\ifdraft
\drafttrue
\ifdraft
\newcommand{\kwnote}[1]{ {\textcolor{orange} { ***Karen: #1 }}}

\else
\newcommand{\kwnote}[1]{}
\fi

\newboolean{anonymous}
\setboolean{anonymous}{false}

\begin{document}
\bstctlcite{limit_authors}

\title{Influence of HW-SW-Co-Design\\ on Quantum Computing Scalability}

\author{
\ifbool{anonymous}{Anonymous author(s)}{
  \IEEEauthorblockN{Hila Safi}
  \IEEEauthorblockA{\textit{Siemens AG, Technology}\\
    \textit{Technical University of}\\
    \textit{Applied Sciences Regensburg} \\
    Munich, Germany \\
    \href{mailto:hila.safi@siemens.com}{hila.safi@siemens.com}}
    \and
  \IEEEauthorblockN{Karen Wintersperger}
  \IEEEauthorblockA{\textit{Siemens AG, Technology}\\
    Munich, Germany\\
    \href{mailto:karen.wintersperger@siemens.com}{karen.wintersperger@siemens.com}}
  \and
  \IEEEauthorblockN{Wolfgang Mauerer}
  \IEEEauthorblockA{\textit{Technical University of}\\
      \textit{Applied Sciences Regensburg} \\
      \textit{Siemens AG, Technology}\\
    Regensburg/Munich, Germany \\
    \href{mailto:wolfgang.mauerer@othr.de}{wolfgang.mauerer@othr.de}}
}
}
\maketitle
\begin{abstract}
The use of quantum processing units (QPUs) promises speed-ups for solving computational
problems. Yet, current devices are limited by the number of qubits and suffer from
significant imperfections, which prevents achieving quantum advantage. To step towards 
practical utility, one approach is to apply hardware-software co-design methods.
This can involve tailoring problem formulations and algorithms to the quantum
execution environment, but also entails the possibility of
adapting physical properties of the QPU to specific applications. In this work,
we follow the latter path,  and investigate how key figures---circuit depth and gate 
count---required to solve four cornerstone NP-complete problems vary with
tailored hardware properties.

Our results reveal that achieving near-optimal performance and properties does not 
necessarily require optimal quantum hardware, but can be satisfied with much simpler 
structures that can potentially be realised for many hardware approaches. Using 
statistical analysis techniques, we additionally identify an underlying general model 
that applies to all subject problems. This suggests that our results may be universally 
applicable to other algorithms and problem domains, and tailored QPUs can find utility 
outside their initially envisaged problem domains. The substantial possible improvements 
nonetheless highlight the importance of QPU tailoring to progress towards practical 
deployment and scalability of quantum software. 
\end{abstract}


\section{Introduction}
NP-Complete problems are of great interest in computer science and
mathematics, as many industrial problems belong to this complexity
class. They are believed to be computationally intractable for
classical computers, at least in the worst case. This means that for
large instances of these problems, it may not be possible to find a
solution in a reasonable amount of time using any known algorithm. 
Industrial use-cases already benefit from
approximating optimisation. These problems can be rewritten as
NP-optimisation (NPO) problems and also include combinatorial 
elements to represent each problem~\cite{sax:20:approximate,bayerstadler:21}. In practice, 
there exist heuristics and approximation algorithms that can be used to 
find good near-optimal solutions to some NP-complete problems by choosing
a trade-off between performance and result quality, albeit it is
known that problems exist that defy such techniques~\cite{VangelisTh2009,10.5555/241938}.

Quantum algorithms in general have the potential to improve both, the quality and
performance of approximate solutions to NP-complete problems~\cite{Fahri_2014}. 
QAOA (Quantum Approximate Optimisation
Algorithm) is a particularly well-known and widely studied quantum algorithm for finding approximate solutions to
combinatorial optimisation problems. 
However, among other factors, current quantum hardware limitations restrict
the potential of using QAOA to solve problems of practical interest.
Quantum computers face different challenges; for instance they are limited to a
relatively small number of qubits, typically ranging from around 50 to 400.
Scaling quantum computers to large numbers of qubits is a difficult
engineering problem that also heavily depends on the specific hardware 
platform. Another problem is that quantum computers are susceptible
to noise and distortions from their environment and suffer from imperfections
in the control signals~\cite{IEEE_Spectrum}, both leading to errors in the
operations performed on the qubits, and limited decoherence times.

Changes to the hardware architecture can influence the connectivity 
between qubits, the coherence time, and the gate error rates. These 
modifications impact the performance and resource requirements of quantum
algorithms, such as the number of gates needed to execute the quantum circuit,
the number of measurements required and the amount of memory and time
needed to store and manipulate quantum states.
In this paper, we consider the effects of such hardware improvements on four NP-
complete problems: Travelling Salesperson (TSP), Number Partitioning (NumPart), 
Maximum Cut (MaxCut) and Maximum 3-Satisfiability (Max3Sat).

This is of particular importance in the current era of noisy, intermediate-scale 
quantum (NISQ) computers, which is expected to last for at least several years
(possibly even decades) until fault-tolerant, perfect quantum computing becomes 
feasible. Yet, there is an increasing interest in utilising NISQ devices in high-
performance computing (HPC) scenarios, and tailoring NISQ devices to problems 
is seen as a possible or even necessary of stepping towards practically
relevant quantum speedups and advantage. As properties of quantum algorithms depend
on QPU (hardware) properties~\cite{wintersperger:22:codes},
hardware-software co-design can help to address some of the key challenges
of current quantum devices~\cite{Li_2021}. By designing algorithms and programs 
optimised for these limitations, it may be possible to surpass 
result quality of more generic approaches. It is also an 
important and promising approach to optimise the performance and
efficiency of quantum computing systems by putting both the 
hardware and software component as a cohesive unit.  The focus 
of this paper is to examine the impact of hardware-software
co-design on quantum computing scalability. We use numerical experiments
to explore the potential for co-design using a hybrid quantum algorithm
(QAOA) applied so several subject problems. The quantum circuits are compiled to 
different types of simulated hardware backends, which are extended
from the topology of the IBM-Q devices. 

The paper is augmented by a reproduction package~\cite{mauerer:22:q-saner},
which is available for \href{https://github.com/quantumdoubleblind/qsw_2023.git}{download} 
(link in PDF).\footnote{We
will place this material on a long-term stable, DOI-compliant location
for the accepted version of this paper.} Some supporting material that
we could not present in the main text is available on the \suppweb.

The rest of this work is structured as follows.
Section~\ref{sec:related} reviews related work. Following in Sec.~\ref{sec:context_foundations}, we explain the principles behind
our approach, and also elaborate on properties of the subject problems.  
Results from numerical experiments, conducted in Sec.~\ref{sec:experiments},
are analysed in Sec.~\ref{sec:evaluation}, which also presents a general
model that universally describes all subject problems. Finally, Sec.~\ref{sec:discussion} discusses the consequences of our findings, 
together with an outlook on future research directions.

\section{Related Work}\label{sec:related}
QAOA has been studied as a promising approach to solve combinatorial
optimisation problems. Several previous works have focused on
optimising QAOA for available quantum hardware. The main challenges
in this context are the limited number of qubits and the high error
rates of current quantum devices. To address these challenges,
different approaches have been proposed, such as the use of
hardware-software co-design, error mitigation techniques~\cite{Temme_2017},
hardware-efficient ansätze~\cite{Fahri_2014} or
hybrid classical-quantum optimisation~\cite{Akshay_2021}. 
Lotshaw~\etal~\cite{Lotshaw_2022} discuss the impact of problem
sizes and complexity on QAOA performance and resource requirements on
contemporary hardware, and also analyse scalability of the algorithm under different
hardware topologies.
Furthermore, Wille~\etal~\cite{Wille:2023} address the challenge
of mapping quantum circuits to the topology of targeted architectures
and present a tool for tackling this problem.
Some works have focused more on designing quantum hardware
that is optimised for specific quantum algorithms. For example in the
work of Kandala~\etal~\cite{Kandala_2017}, a superconducting quantum
processor was optimised for the variational quantum eigensolver (VQE)
algorithm. The paper shows that this approach can significantly
reduce the number of gates required to implement VQE on the
hardware. In~\cite{Li_2021}, an architecture design-flow for
superconducting quantum computers is proposed that finds a trade-off
between optimisation of the processor's yield rate and a mapping with minimal
gate-overhead. The ansatz is compared to other designs using the example
of VQE for quantum chemistry calculations.
Furthermore, Linke~\etal~\cite{Linke_2017} assert that
co-designing quantum applications for specific purposes is crucial to 
successfully utilise quantum computers in the near future. 
They reach this conclusion by comparing identical quantum algorithms on 
two different hardware platforms.

\section{Context and Foundations}\label{sec:context_foundations}
In the following, we lay some foundations necessary to
understand our approach and rationale behind the experiments..

\subsection{Quantum Optimisation with QAOA}
QAOA is a widely used variational hybrid
quantum algorithm on NISQ hardware developed by
Fahri ~\etal in 2014~\cite{Fahri_2014}.
The algorithm has shown promising results on small-scale 
quantum devices for several optimisation problems, including
the MaxCut problem, TSP~\cite{Shaydulin_2019}, or similar
problems~\cite{Feld:2018}.
As quantum hardware continues to improve, QAOA and other
quantum optimisation algorithms are expected to play
an increasingly important role in solving real-world problems.

\subsubsection{Algorithm}
QAOA produces approximate solutions for combinatorial optimisation
problems, which are described by a problem Hamiltonian \(H_p\). 
The algorithm consists of several layers of parameterised unitary
operators \(U(\beta, \gamma)\). As the number of layers \(p \geq 1\),
\(p \in \mathbb{N}\) increases, the quality of the approximation
improves~\cite{Fahri_2014}. First, the quantum register is 
initialised in a well-defined state and after applying the unitary
operators, the expectation value of \(H_p\) is measured in the final state.
The parameters \(\beta, \gamma\) of the quantum circuit are
optimised by classical methods such that the expectation
value of \(H_p\) is minimised. 

Each layer consists of two different kinds of unitaries, \(U(\beta_i)
= e^{i\beta H_B}\) and \(U(\gamma_i) = e^{i\gamma H_P}\). The algorithm
applies a mixer Hamiltonian, typically a Pauli-X operator, to
each qubit using the \(U(\beta_i) = e^{i\beta H_B}\) unitary. This
is followed by a combination of single qubit \(Z-\)rotations
\(R_Z(\gamma_i)\) and two-qubit rotation gates \(R_{ZZ}(\gamma_i)\)
composing the \(U(\gamma_i) = e^{i\gamma H_P}\) unitary. Multiple
layers of this process correspond to a discretized time evolution
governed by the Hamiltonians \(H_P\) and \(H_B\). The algorithm's
initial state is usually the ground state of \(H_B\), prepared
using Hadamard gates \(H\). To optimise the objective function,
the quantum circuit is executed multiple times, and the qubits
are measured in the computational basis. The mean of the expectation
values of \(H_P\) for each measurement outcome is minimised by the 
classical optimiser, and the optimal solution is derived as the
state or bit string with the lowest energy expectation value in the
probability distribution obtained from the final set of parameters.
The algorithm determines the minimum value of the objective function
specified in quadratic unconstrained binary form (QUBO). A classical
algorithm that can efficiently sample the output distribution of
QAOA even for \(p = 1\), cannot exist based on reasonable
complexity-theoretic assumptions. This indicates the possibility
of quantum advantage, but practical utility on real-world problems
require further investigations~\cite{wintersperger:22:codes}.

\subsection{Translation of algorithms to quantum hardware}

When programming a quantum algorithm, initially no restrictions on the
type of gates being used or the interaction between qubits is made.
However, to execute a certain quantum algorithm on a specific hardware
backend, it needs to be compiled~\cite{venturelli2019quantum} to the properties of the backend,
which is also called \emph{transpilation}~\cite{qiskit_transpiler}. The most important
properties of a quantum computer that influence the transpilation of
circuits are the size of the backend, that is, the number of qubits
available, their geometric arrangement and connectivity, and the
native gate set. For the actual execution of the circuit, also other
factors such as the fidelities of gate operations, initialization and
measurement as well as the decoherence and gate operation times play
an important role.

The connectivity measures the number of other qubits one qubit can
interact with, and thus the ability to perform a two-qubit gate
operation between them. If a two-qubit gate needs to be executed
between qubits which are not connected, a \sgate can
be applied to swap the states of two qubits. The geometric layout and
connectivity of the QPU can be depicted by a graph with nodes
representing the qubits and edges connecting two qubits if an
interaction between them is possible.  Analogously, the circuit that
is executed can also be represented by a graph, which has an edge
between two nodes, if a two-qubit gate is performed between the
corresponding qubits.  The transpilation process maps this circuit
graph to the hardware graph, while taking into account further
restrictions, such as the native gate set.

Due to the decomposition of gates into the native gate set of the
hardware as well as the insertion and further decomposition of
\sgates, the transpiled circuit contains more gates in total. This is
crucial in the current NISQ-era, as each gate introduces an error, and
thus the quality of the results is expected to drop with a growing
number of operations. Moreover, the circuit depth is increased, which
measures the maximum length of the circuit accounting for parallel
execution of gates, as well as the overall runtime of the algorithm,
due to the finite execution time of each gate. The available quantum
computers only have a limited decoherence time, in which operations can
be performed, that should not be exceeded by the algorithm runtime.

Thus, to increase the performance of quantum algorithms on near-term
quantum computers, the number of gate operations should be
minimised. This could be achieved by designing optimised algorithms or
by increasing the connectivity of the hardware.

The ability to modify these properties depends on the type of the QPU
being used.  Today, several different types of quantum computing
hardware exist, which differ by the physical implementation of
qubits. The two states \(\ket{0}\) and \(\ket{1}\) can be encoded in
various different ways such as the naturally occurring discrete energy
levels of single ions or atoms, the effective energies 
of superconducting circuit elements or in the spatial modes of single
photons, to just name a few~\cite{Bruzewicz_2019, 
henriet_quantum_2020, bravyi_future_2022, Slussarenko_2019}. Along with the
choice of qubits, also the control and readout techniques, the
infrastructure requirements (\eg, if cooling with a cryostat is
needed) and the properties relevant for mapping between logical and
physical circuits, such as the number of qubits, the native gate set
and the connectivity are different for each type of QPU. This means
that the performance of a quantum algorithm usually heavily depends on
the type of hardware that it is running on.

\subsection{Problem selection}
In this paper, we focus on problems that belong to complexity class
\CC{NP}-Complete (\CC{NPC}), as it contains practically relevant problems that,
assuming the usually uncontended \(\CC{P}\neq\CC{NP}\) hypothesis,
cannot be efficiently solved on a classical machine, and are in most
instances also hard to approximate, as is textbook
knowledge~\cite{books/daglib/0030297}.

A decision problem \(p\) is in NPC if a solution can be determined by
a non-deterministic Turing machine in polynomial time (\ie,
\(p\in\text{NP}\)), and is additionally NP-hard, which means that any
other problem in NP can be reduced to \(p\) in polynomial time. We
investigate the fundamental MaxCut, NumPart, TSP and Max3Sat problem.

\subsubsection{Maximum Cut}

Given an undirected graph \(G = (V, E)\) composed of vertices \(V\) and the 
set of edges \(E\), the objective is to partition the vertices into two 
disjoint sets, S and T, while maximising the number of edges that cross the 
partition:

\begin{equation}
\max_{x_{i,j}} \sum_{(i,j) \in E} (2 x_i x_j - x_i - x_j),
\end{equation}

 where \(x_{i}\) is a binary variable that takes the value
 \(1\) if vertex \(V_{i}\) lies in the first subset S and
 \(0\) if it lies in the second subset T.

\subsubsection{Number Partitioning} 
Let \(x_{1}, x_{2}\, \ldots, x_{n}\) be a set of positive
integers. The objective is to divide the set into two subsets S and T,
while minimising the difference between the sums of the two non-empty
subsets
 \begin{equation}
\min_{x_{i}} \left(\sum_{i=1}^{n} a_i x_i - 
                    \sum_{i=1}^{n} a_i (1- x_i)\right)^{2},
 \end{equation}

 where \(x_i = 1\) if \(a_i\) is assigned to subset S and \(x_i = 0\)
 if \(a_i\) is assigned to subset T. Note that we minimise the square
 of the expression, since a QUBO formulation is not able to represent the
 alternative of absolute values.

\subsubsection{Travelling Salesperson} 
Given a set of \(n\) cities \(1,2,\ldots, n\), the travelling
salesperson problem determines the shortest path, whilst starting and
ending at the same city and visiting each location exactly once

\begin{equation}
\min_{x_{i,j}} \sum_{i=1}^{n} \sum_{j \neq i, j = 1}^{n} c_{i,j} x_{i,j},
 \end{equation}

 where \(x_{i,j} = 1\) if the path goes from city \(i\) to city \(j\)
 and \(x_{i,j} = 0\) otherwise.

\subsubsection{Maximum 3-Satisfiability} 
Given a set of m clauses \(1,2,\ldots, m\) , each consisting of three
Boolean variables or their negations, Max3Sat seeks
to find an assignment of truth values to the variables that satisfies
the maximum number of clauses. The objective function can be expressed
as follows:

\begin{equation}
\max_{x_{i,j}} \frac{1}{m} \sum_{i=1}^{m} w_i \sum_{j=1}^{3} x_{i,j},
\end{equation}

where \(w_{i}\) is the weight of clause \(i\), \(c_{i}\) is the number
of literals in clause \(i\) that are satisfied by the assignment, and
\(m\) is the total number of clauses.

We selected this set of problems for two reasons, one of which is that
they represent significant industrial use cases associated with
them. MaxCut has various industrial applications in network
optimisation and clustering. Amongst other things it is used for
targeted advertising, recommendation systems as well as identifying
the ideal placement, for instance, for hospitals or subway stations to extend and
improve infrastructures. NumPart can be used for load
balancing. The goal could be to divide a set of tasks among machines
in a way that minimises the differences in workload. It can also help
to find an optimal division of orders among workers. The TSP is
well known and is commonly
applied in the logistics and transportation industry~\cite{Alridha_2021}.
Just as importantly, Max3Sat is
used in the design of digital circuits, where the goal is to minimise
the number of gates needed to implement a logic function, thus reducing the
overall complexity of the circuit. This can help save costs and leads
to a better performance. It is worth noting that many other
NP-Complete problems have similar applications in various industrial
settings.  

The other reason to chose this set of problems is that decision
problems are less common in industrial use-cases than approximate
optimisation problems. For example, in the case of the travelling
salesperson one could ask "what are possible short routes".
Accepting for small deviations from optimal solutions can lead 
to significant savings in time and effort
for many problems, which usually is a more desirable outcome in
practical applications~\cite{sax:20:approximate}.  This particular
problem set contains problems from three different complexity 
classes when described as NP (nondeterministic polynomial time) 
problems---APX-Complete, NPO-Complete and MAX-SNP. This helps
us compare the scalability within the same complexity class,
as well as across different complexity classes.

\subsection{Complexity classes in NP optimisation problems}
APX-complete problems are considered to be the hardest problems to
approximate within a constant factor in polynomial time, assuming
\(\CC{P} \neq \CC{NP}\). MaxCut belongs to this complexity class.
NPO-complete problems include the TSP and
NumPart problem. These problems are characterised by the task of
finding an optimal solution that satisfies a set of constraints, and
are at least as hard as the hardest decision problems in NP. Unlike
APX-complete problems, NPO-complete problems may not have a
constant-factor approximation algorithm that runs in polynomial
time. On the other hand, MAX-SNP consists of optimisation problems that
can be expressed as a Boolean formula in conjunctive normal form,
where each clause is a disjunction of at most k literals. The main
difference between these complexity classes lies in the 
available approximation algorithms~\cite{Crescenzi_1994}.

\begin{figure*}[htbp]
    \includegraphics{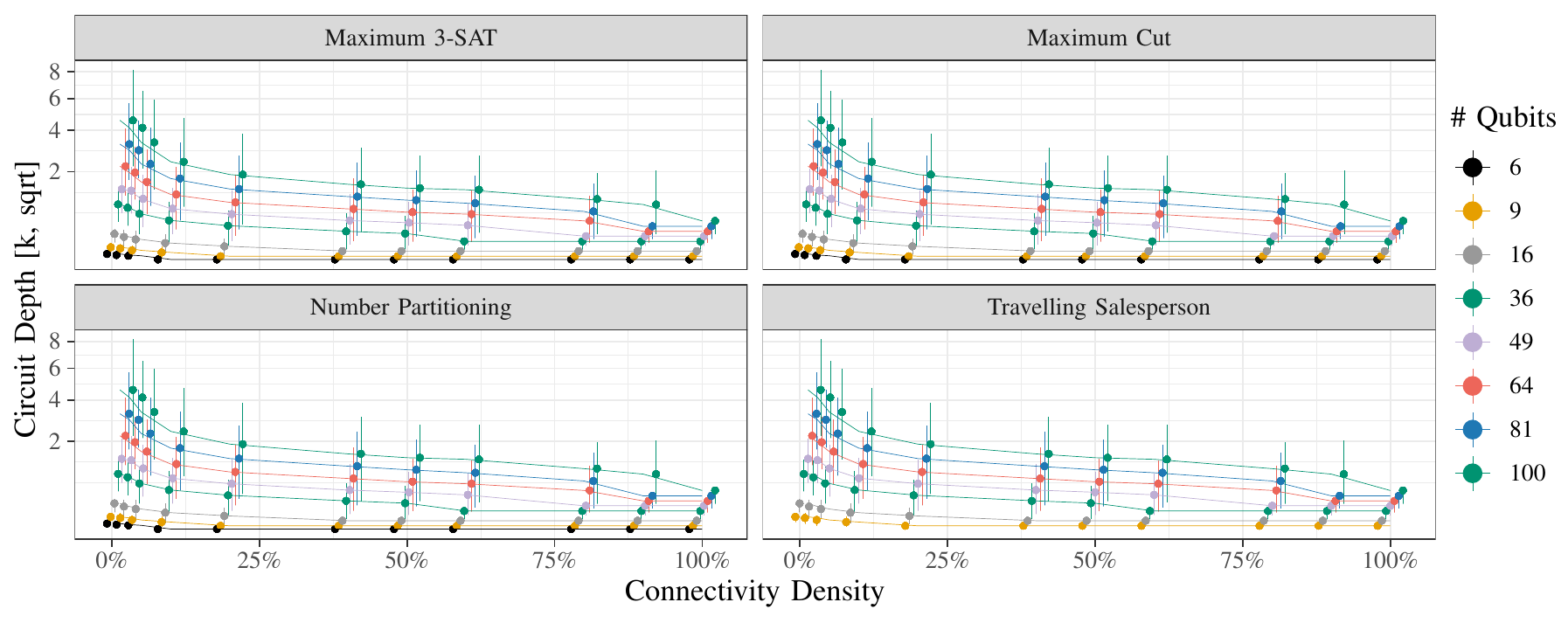}\vspace*{-1em}
    \caption{Span of achievable circuit depths for the subject problems in different
    sizes, plotted over varying degrees of connectivity. Data points for each density
    are slightly displaced horizontally to make point ranges visible; connecting
    lines do not provide a fit to the data, but are only used to guide the eye.}%
    \label{fig:depth_over_density}
\end{figure*}

\section{Experiments}\label{sec:experiments}
\subsection{Setup} In this work, the quantum circuits are designed and transpiled
using Qiskit~\cite{Qiskit}. We study the QAOA-circuits for several
instances of the four problems described above after
being transpiled to hardware backends with different properties. As a
starting point, a backend with 127 qubits is chosen that matches the
geometric layout and connectivity of the current IBM-Q devices, the
so-called \emph{heavy-hex}-geometry~\cite{ibmq_heavy_hex}. The native
gate set corresponds to that of IBM-Q, containing the following gates:
Rotation $RZ$, phase shift $SX$, Pauli (Not) $X$, and controlled X
(\CXgate). In principle, also the influence of noise on the transpilation
process could be modelled in Qiskit, which is, however, not in the
scope of this work.

We investigate the depth and number of \sgates of the
transpiled circuits depending on the connectivity and size of the
backend. The circuit depth measures the overall length of the circuit,
taking into account also parallel execution of gates. The
\sgate counts are derived by mapping each circuit a
second time using an extended native gate set including the
\sgate. This prevents the latter from being decomposed
into other gates.

The connectivity of the backend is measured in terms of
a \emph{connectivity density}
\begin{equation}
c = \frac{N_C}{N_{C,\mathrm{max}}},
 \end{equation}
with $N_C$ denoting the total number of edges in the hardware
graph and $N_{C,\mathrm{max}}=N(N - 1)/2$ the maximal number of
edges for $N$ qubits. 
While $c=1$ describes a device with all-to-all connectivity such as in ideal
simulations, the heavy-hex-geometry has a connectivity density of
$c\approx 1.8\%$. This value corresponds to each qubit having on average
$2.27$ nearest neighbours. 
In the experiments presented below, the connectivity density is increased
by randomly adding connections between pairs of qubits until the desired
value is reached. The average number of nearest neighbours per qubit
grows linearly with the connectivity density.

\subsection{Problem Mapping}
All problems in NP can be reduced to Quadratic Unconstrained Binary
Optimisation (QUBO) problems. The QUBO formulations in this work follows 
the Ising formulations given by Lucas~\cite{Lucas2014}.

\subsubsection{Maximum Cut}
MaxCut can be cast using binary variables \(x_i\), where \(x_i = 1\)
indicates that node \(i\) belongs to the first subset, and \(x_i = 0\)
indicates that it belongs to the second subset. If an edge connecting
nodes \(i\) and \(j\) is part of the cut, then one of \(x_i\) 
and \(x_j\) is equal to zero and the other one is equal to
one, resulting in \(H_{ij} = (x_i + x_j - 2x_ix_j)\) being \(1\),
whereas \(H_{ij}\) equals \(0\) if \(x_i = x_j\). The goal is to find
the MaxCut by maximising the sum of \(H_{ij}\) over all edges of
the graph or in other words minimising the sum over \(-H_{ij}\).
The optimal solution is the ground state of the Hamiltonian 
\begin{equation}
H_P = \sum_{(i,j) \in E}(2x_i x_j - x_i - x_j)   
\end{equation}
which serves as the objective function for the QAOA algorithm 
to find the minimum solution.

\textbf{Setup:}
The MaxCut problem graphs \(G = (V, E)\) were characterised 
by their number of nodes \(N = |V|\), and the graph density
\(d\), defined as the ratio of the number of edges \(|E|\) to
the maximum possible number of edges \(|E_{\mathrm{max}}|\) 
in a clique comprising \(|V|\) nodes. The value of \(d\) ranges
from \(0\) to \(1\) and is set to \(d=0.7\) for the experiments in this work. Each node in the graph is represented by one qubit, so the problem size given in numbers of qubits directly corresponds to the number of nodes.

\subsubsection{Number Partitioning}
NumPart can be reformulated as a QUBO problem using binary 
variables \(x_i\) where \(x_i = 1\) indicates that \(a_i\) 
belongs to subset S and if \(x_i = 0\), \(a_i\) belongs to
subset T. The objective is given by the Hamiltonian
\begin{equation}
H_P = \Big(\sum_{i=1}^{n} a_ix_i\Big)^{2} - \Big(\sum_{i=1}^{n} a_i\Big)^{2}   
\end{equation}
which represents the difference between the sums of S and T.
The goal is to find the minimum value of the Hamiltonian, 
which corresponds to the optimal partitioning of the set.

\textbf{Setup:} The NumPart problem was generated as a list of length
\(n \in \mathbb{N}\). Each number was generated randomly,
and the corresponding field index is represented by one qubit.

\subsubsection{Travelling Salesperson}
The objective function of the TSP, describing the total length of 
the tour, is given by the following Hamiltonian:
\begin{equation}
H_C = \sum_{i,j \in E,i\neq j}^{N} c_{ij} \sum_{k=1}^{N} x_{i,k} x_{j,k+1}     
\end{equation}
where \(c_{i,j}\) is the distance between nodes \(i\) and \(j\);
\(x_{i,k}\) is a binary variable that is equal to \(1\) 
if node \(i\) is visited at position \(k\) in the tour (and 0 otherwise); 
and \(N\) denotes the total number of nodes in the TSP instance. 
The sum over \(k\) enforces the ordering of the nodes in the tour. Note that
\(x_{i,N+1}\) is equivalent to \(x_{i,1}\), so the tour loops
back to the starting node and in our case \(c_{i,j} = c_{j,i}\).
To ensure that each city is visited exactly once in the tour and that at each position in the tour there is exactly one city, the corresponding penalty terms are added to comprise the final Hamiltonian:

\begin{equation}
H = A \sum_{j=1}^{n}(1-\sum_{k=1}^{N} x_{j,k})^{2}
+ A \sum_{k=1}^{n}(1-\sum_{j=1}^{N} x_{j,k})^{2}) + H_C, 
\end{equation}
where  \(A\) controls the strength of the penalty.

\textbf{Setup:}
The TSP problem graphs were represented as a complete, 
undirected graph (so \(c_{i,j}=c_{j,i}\)), where the nodes represent the cities and the
edges represent the distances between them. A randomly
generated distance matrix determines the distances between
the cities. For \(N\) nodes, we have \(N^2\) qubits and a \(N \times N\) 
distance matrix. The graph density for TSP
is defined analogously to MaxCut and also set to \(d=0.7\).

\subsubsection{Maximum 3-Satisfiability}
To formulate this problem as a QUBO, we introduced three
binary variables \(y_i\), \(y_j\)
and \(y_k\), where each variable can take the value \(0\) or \(1\).
The Hamiltonian for this problem is given by
\begin{equation}
H = \sum_{i=1}^m (1-\frac{1}{2} (y_i + y_j + y_k - 1)   
\end{equation}
where m is the number of clauses. The Hamiltonian is
minimised when the number of satisfied clauses is maximised.

\textbf{Setup:}
The values for each variable were generated randomly. The number of
clauses added to the number of variables equals the number of qubits
necessary to represent our QUBO formulation. In our experiments we
use a fixed number of \(3\) variables per clause. According to
Figure~\ref{fig:sat_instances_depth}, the ratio between the number
of clauses and number of variables has little to no effect on the
circuit depth, suggesting it can be disregarded in the experiments.

\subsection{Circuit Mapping} The hardware
backend is described by a connectivity graph given in the form of
tuples and a native gate set. The transpilation process in Qiskit
consists of several steps: First, the circuit is optimised, for
instance by combining several single-qubit gates into a single
one. Then, all gates which do not belong to the native gate set, such
as gates with more than 2 qubits, are decomposed into the native gate
set. The next step is to find an optimal placement of the logical
qubits in the circuit to the physical qubits of the hardware, which
corresponds to a direct mapping of the problem (or algorithm) graph to
the hardware graph. Thereby, \sgates are inserted, if necessary, and
the mapping is determined to minimise the number of \sgates. For the
circuits presented here, the standard mapping method of Qiskit has
been used, which includes a stochastic placement of \sgates. 
After the mapping, the inserted
\sgates are being translated into the native gate set (if necessary),
and the circuit is optimised once more, accounting \eg, for possible
concatenations of gates.

The aggressiveness of depth optimisation varies
between four levels~\cite{qiskit_transpiler}
(level \(n\) includes all measures of levels \(k<n\)):

\begin{itemize}
\item $0$ (off): Map without optimisation.
\item $1$ (light): Collapse adjacent gates that cancel each other.
\item $2$ (medium): Noise-adaptive layout, gate cancellation based 
on gate commutation relationships.
\item $3$ (heavy): Replace blocks of gates with (different, yet 
semantically equivalent) optimised gate sequences.
\end{itemize}

Our numerical experiments were performed at optimisation level 2, which provides
a good trade-off between result optimality and required computational effort.
This choice is further justified in Section~\ref{sec:threats}.

\section{Evaluation}\label{sec:evaluation}
We commence to discuss the outcomes of our numerical experiments
in the following, and then find common patterns in the data using
statistical analysis techniques.

\subsection{Numerical Results}

\subsubsection{Circuit Depth and \sgate Count}
The depth of quantum circuits is analogue to classical
runtime---the more gates are involved in a circuit, the longer 
a quantum computation takes---, but also key to understanding the
capabilities of NISQ machines, as increasingly deep circuits are 
subject to growing amounts of noise and decoherence, eventually 
leading to entirely stochastic results that do not provide information
about the problem at hand. Recall that quantum circuits generated for 
specific problem formulations are produced by a uniform generation 
mechanism, but vary with instance size and characteristics of the 
individual instances.

\begin{figure*}[htbp]
    \includegraphics{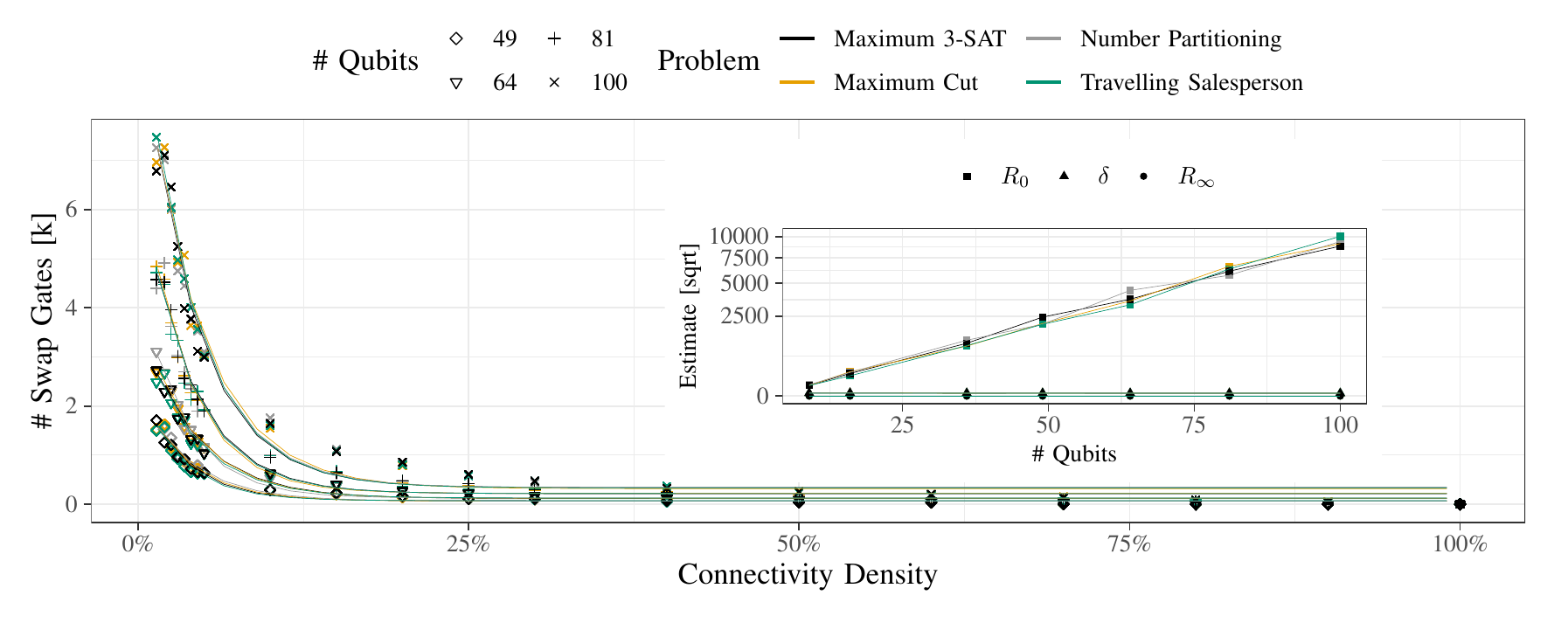}\vspace*{-1.5em}
    \caption{Outer: Empirical observation of \sgate count 
    decrease with increasing \cd (points),
    together with fits obtained by the negative exponential model
    (solid lines) for the subject problems. Inset: Corresponding 
    model coefficients.}%
    \label{fig:swap_scaling_density}
\end{figure*}

As one of our goals is to understand the effects of varying
degrees of connectivity in (hypothetical) QPUs, first consider 
Fig.~\ref{fig:depth_over_density}, which shows the achievable circuit 
depths for a given degree of (extended) connectivity
for the subject problems in various instance sizes given by 
the amount of required qubits. Since mapping between 
logical and physical circuits is performed by a stochastic algorithm, we
obtain a range of depths for varying connectivity densities and
qubit count. Data points in Figure~\ref{fig:depth_over_density}
represent mean values over 20 compilation runs. It is immediately
apparent that even small increases in circuit depth over the base
connectivity of IBMQ's heavy hex topology lead to considerable reduction
of the circuit depth in a similar way for all of the problem types
considered here. Likewise result
variability increases considerably towards smaller degrees of connectivity. Both, strength of
variability and circuit depth, converge for densities exceeding 25\%.

Figure~\ref{fig:swap_scaling_density} shows the amount of \sgates
that are required for a given connectivity density (we address the
inset in Sec.~\ref{sec:modelling} below). Since \sgates
are necessary to bring qubits into physical adjacency when multi-qubits
operations must be applied on topologically not adjacent qubits, they
can be seen as overhead that arises from restricted connectivity 
densities. As the figure shows, zero \sgates are required when the
density reaches 1.0, as the need to logically move qubits by swapping
them around in the circuit does not arise in this case. Similar to
circuit depth, we can observe a steep decline in \sgate count with
increasing connectivity density, and a plateauing of the count form
densities of about \(30\%\) onwards.

\subsection{Statistical Modelling}\label{sec:modelling} 

\begin{figure}[htbp]
    \includegraphics{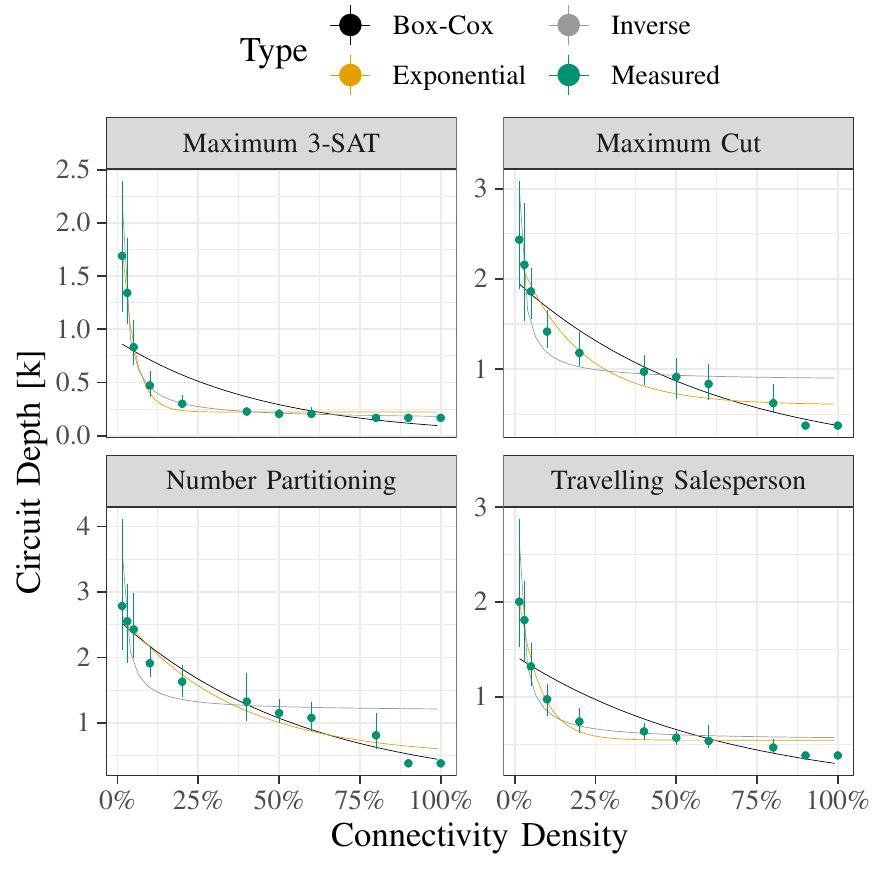}\vspace*{-1em}
    \caption{Comparing models (inverse, Box-Cox, negative 
    exponential; solid lines) against the  empirically 
    measured data (points). The graph shows problem 
    instances each requiring 64 qubits; graphs for other 
    sizes exhibit similar characteristics and are available
    on the \suppweb.}%
    \label{fig:density_fits}
\end{figure}

While it is obvious from Figures~\ref{fig:depth_over_density} 
and~\ref{fig:swap_scaling_density} that even slightly improved connectivity
density results in substantial reductions on circuit depth independent of the
specific problem, it is pertinent to further characterise this empirical
observation. To find simple models that accurately describe the observed
phenomena, we fit statistical models to the available data.

The sharp decrease of circuit depth with increasing \cd, modelled in
general by a functional dependency of the form \(d(\varrho) =
f_{\text{P}}(\varrho)\) (where \(\text{P}\) denotes specialisation
for a specific problem) suggests an inverse
(\(f(\varrho) \sim 1/\varrho\)) or negative exponential
(\(f(\varrho) \sim \exp(-\varrho)\)) relationship. The empirical
results for these ansätze (linear univariate regression~\cite{Fahrmeir:2013}
for the inverse relationship,
non-linear regression~\cite{Bates:2009} for the negative exponential), 
together with a linear regression fit based on a Box-Cox 
transformation~\cite{Venables:2002} of the data,\footnote{The transformation uses
a maximum-likelihood estimate to determine an optimal non-linear transformation
to minimise the standard deviation of regression residuals, which could suggest
desirable other forms of functional
dependencies than the two considered variants.}
is shown in Fig.~\ref{fig:density_fits}.
Visually, it is obvious that the linear regression based 
approaches result in a sub-optimal match between model and data, whereas the 
negative exponential ansatz
\begin{equation}
d(\varrho) = R_{\infty}+(R_{0}-R_{\infty})\cdot
e^{-\exp(\delta)\cdot\varrho}\label{eq:nls_model}
\end{equation}
describes the data very well.\footnote{A straightforwards logarithmic
transformation of the data, which would allow us to deploy a simpler
linear univariate regression model, does not produce satisfactory results;
while the variation of the decay constant is small
across instance sizes for each subject problem, it is nonetheless large
enough to warrant different bases for each log transformation, which would
need to be estimated in a prior modelling step.}
Based on the data for each subject problem, parameters
\(R_{0}\), \(R_{\infty}\) and \(\delta\) are obtained for varying \cds.
\(R_{\infty}\) denotes the horizontal asymptote towards large values of the
\cd, \(R_{0}\) is the  (extrapolated) circuit depth estimate for vanishing
\cd at \(\varrho=0\), offset by \(R_{\infty}\).  Coefficient
\(\delta\) represents the natural logarithm of the exponential rate constant,
and characterises the speed of decline with increasing \cd. 

\begin{figure}[htbp]
    \includegraphics{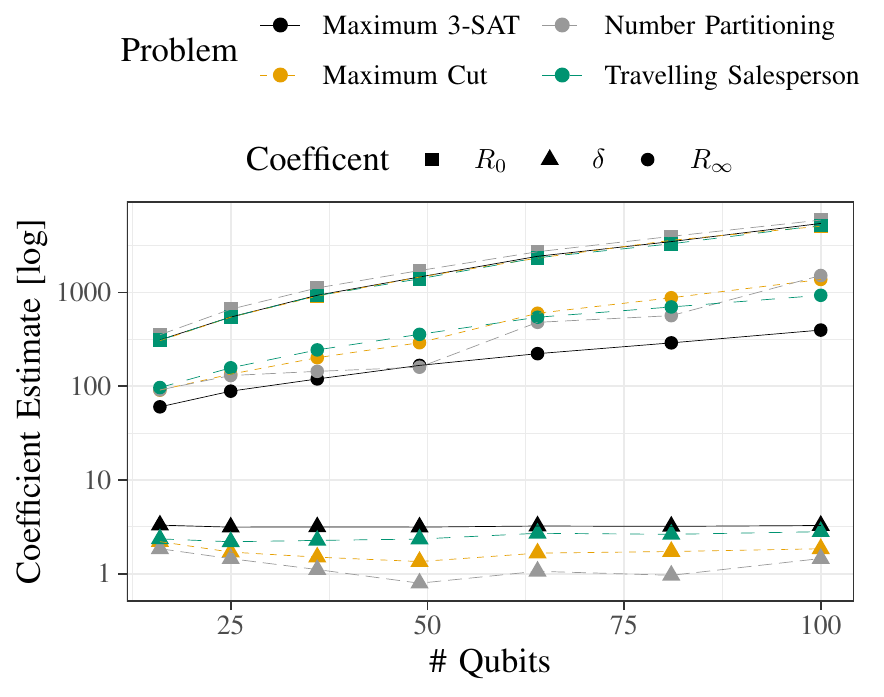}\vspace*{-1em}
    \caption{Coefficients for the non-linear negative exponential fit
    described by Eq.~\ref{eq:nls_model} to the circuit depths for all subject
    problems over varying instance sizes. Connecting lines
    have no significance, and are only used to guide 
    the eye.}\label{fig:density_model_coefficients}
\end{figure}

Consider Figure~\ref{fig:density_model_coefficients}, which summarises the
evolution of model parameters for the circuit depth with increasing instance sizes for all subject 
problems. Note that for each problem and instance size, we numerically
compute circuit depths for a range of \cds, and then fit Eq.~\ref{eq:nls_model}
to the data. Consequently, each combination of problem and instance
size delivers \emph{three} parameters. The evolution of these parameters
with increasing instance sizes is shown in the figure.

Apart from some smaller variations for NumPart, the rate
constant \(\delta\) is stable across instance sizes---that
means that exponential improvements in circuit depth with
increasing \cd are achieved nearly uniformly across the
full spectrum of instance sizes. As gains are mostly
independent of the problem, we hypothesise that this
behaviour holds as a general law for QAOA-based approaches.
Circuit depths in the limits of zero and full connectivity,
obviously increase with increasing problem size. 
It is however important to observe that the evolution is also
very similar across subject problems, again hinting at a general
property of QAOA circuits.

The inset in Fig.~\ref{fig:swap_scaling_density} can be 
interpreted similarly, except that we use \sgate counts 
instead of circuit density as dependent quantity for the
model in Eq.~\ref{eq:nls_model}. Since it is an a-priori
invariant that the \sgate count needs to reach zero for full 
connectivity (qubits do not need to be swapped around if
interactions between any possible pair can be implemented
natively), we fit a restricted form of Eq.~\ref{eq:nls_model}
where the asymptote \(R_{\infty}\) is constrained to vanish.
The obtained parameters show even better agreement across
subject problems than for the circuit depth, which can
be explained by the fact that \sgates constitute
\enquote{overhead} gates to compensate for connectivity
deficiencies. As our results show, this impacts all problems
equally. Yet, the observed exponential decrease with
increasing \cd underlines that even small 
changes have substantial impact on QC utility.

\subsection{Implications for Co-Design}

In the previous section we have seen that the circuit depth as well as
the number of inserted \sgates is already reduced by a
significant amount when increasing the connectivity density to
intermediate values of about $30\%$. This value increases
slightly with the problem size, but does not depend on the problem
type, as shown in Fig.~\ref{fig:density_model_coefficients}.
Overall, we can state that full connectivity is not essential
to decrease the resource requirements for the considered QAOA circuits.

In general, a quantum computing device with connectivity density
between $10\%$ to $50\%$ would be an appropriate choice for all of the
four problem types. The \sgate overhead might be reduced
further, if also the geometric layout of the hardware graph directly
matches that of the problem graph, opposed to randomly adding
connections as done in this work. A connectivity density of $c=10\%$
corresponds to each qubit being connected to $15$ nearest neighbours
on average, for $c=50\%$ this increases to $64$ neighbours. Apart from
the reduced number of gates in the circuit, a higher qubit
connectivity is desirable for implementing efficient
error-correction schemes which in turn require a lower overhead in the
number of physical qubits needed to encode a logical qubit, such as
low-density parity check (LDPC) codes~\cite{breuckmann_quantum_2021,
  bravyi_future_2022}.

The connectivity of currently available quantum computers depends on
the type of quantum hardware being used. Architectures based on
superconducting qubits, such as the devices built by IBM and Google,
are currently limited to nearest neighbour connectivity (so
$c\approx 3.2\%$)~\cite{stassi_scalable_2020, Linke_2017,
  acharya_suppressing_2023}. There exist several ideas to increase the
connectivity. One common approach is to couple several qubits to a
quantum bus, either directly via tuning their frequencies in and out
of resonance with the bus~\cite{Song_2019}, or indirectly via
additional flux qubits~\cite{stassi_scalable_2020}, which are variably
tuned. The latter architecture has the advantage of lower cross-talk
and longer coherence times, since the data qubits can be operated at
their optimal frequencies. Other ideas include using sparse
connections but with non-trivial topologies, extending the
architecture to 3D or using long cables to connect distant
qubits~\cite{bravyi_future_2022}.  With the quantum bus setup proposed
in~\cite{stassi_scalable_2020}, the connectivity could theoretically
be increased such that two-qubit-gates can be performed between all
pairs of qubits, superseding the insertion of
\sgates at all. 
Nevertheless, realising such a setup with only an
intermediate connectivity, as suggested by our findings, will
in any case benefit the practical implementation.

In general, increasing the number of
connections between qubits can lead to a higher probability for
crosstalk. This term describes unwanted interaction between qubits or
between qubits and the control signals, which means that a gate pulse
can effect other than the target qubit(s) or local gate operations are
disturbed by other gate operations applied in parallel. These effects
are especially detrimetal for implementing error correction, which
assumes that gate errors only affect the state of the target qubits.
For superconducting qubits, crosstalk can be reduced by using qubits
with tunable frequencies (see~\cite{stassi_scalable_2020}) and / or
tunable couplers to switch connections dynamically on and off. In
addition, optimizing the pattern of tunable qubit frequencies and gate
schedules via software can also lead to substantial
improvements~\cite{ding_systematic_2020}. On the other hand, architectures with fixed-frequency qubits and fixed
couplers like IBM-Q that do not allow for such optimisation suffer from fewer sources of noise. In this case, 
optimised gate schedules are being used to minimise 
crosstalk~\cite{murali_software_2020}.

Quantum computers based on cold neutral atoms and Rydberg-interactions
already feature a higher connectivity of about $1$:$10$ to $1$:$20$ in
2D- and 3D-layouts~\cite{henriet_quantum_2020}, which would
correspond to $c\approx 8$-$16\%$ for the heavy-hex-based layout.
The connectivity can be further increased by using
higher energy levels for the Rydberg interaction, which, however,
might lead to a higher susceptibility to noise and become technically
challenging. Another approach is shuttling of atoms during the
computation to allow for two-qubit-gates between arbitrary pairs of
qubits~\cite{bluvstein_quantum_2022}. In general, crosstalk is quite
low for neutral atom qubits~\cite{Xia_2015, Graham_2019}, since their
distances can be made large enough to avoid unwanted excitations of
spectator qubits. Also, increasing the qubit connectivity is not necessarily
related to higher crosstalk for this platform.

In contrast to the two previous examples, trapped ion quantum
computers are characterized by an all-to-all connectivity, which means
that two-qubit gates can be performed between each pair of qubits, but
also up to $20$ qubits can be entangled~\cite{friis_2018}. On the
other hand, ion trap setups are more difficult to scale to larger
numbers of qubits. The most common technique stores ions in a linear
string and is limited to qubits numbers in the range of $50$. 
This has to be compared to superconducting qubits and neutral
atoms, which currently offer up to $\approx 400$~\cite{ibm_433qubits} and
$\approx 100$~\cite{henriet_quantum_2020, bluvstein_quantum_2022} qubits,
respectively, and, in the latter case, are also easier to scale. Ion
strings with larger number of ions are expected to suffer from lower
gate speeds, higher crosstalk and background
noise~\cite{brown_co-designing_2016}. To realise trapped-ion devices
with larger number of qubits, mainly two different approaches exist:
coupling several linear traps via photonic interconnects or shuttling
of ions in a 2D trap. While the first approach is more simple to
realise, it is affected by higher crosstalk due to residual illumination
of ions which are not targeted by a gate operation. Crosstalk can be
reduced by careful design of pulse sequences or improved laser
focusing, as well as by using refocusing
schemes~\cite{parrado-rodriguez_crosstalk_2021}. While the first
option can be broadly attributed to the software domain, the latter
two options are deeply intertwined with the core physical realisation
of QPUs.

Finally, to find the most suitable platform for a quantum program,
a trade-off between several properties such as the connectivity, number
of qubits or error-rates has to be made. 

\section{Threats to Validity} \label{sec:threats}

\subsection{External Validity}
Our scope is limited to the Qiskit compiler and the base topology of
IBM-Q devices. It is important to note that using different compilers
may result in varying circuit properties (see Salm~\etal~\cite{Salm:2021}),
which means that our findings may not be applicable to other compilers.
Additionally, different topologies may yield different outcomes.
Furthermore, we only consider QAOA. There are other (variational) quantum
algorithms as well as different types of the QAOA algorithms, that tackle
NP optimisation problems~\cite{Cerezo_2021}. While it is possible to model
the impact of noise on the transpilation process in Qiskit, which would
effect the circuit depth, it falls outside the scope of this
work~\cite{Paler:2021}.

\subsection{Internal Validity}
Our observations rely on controlled numerical experiments that depend on explicit parameters, but may also be influenced by confounding factors.
In the following, we consider various
possible confounding factors, and find that they pose moderate to 
no risk to the validity of our study.

\subsubsection{Influence of Mapping Optimisation}
\begin{figure}[htbp]
  \includegraphics{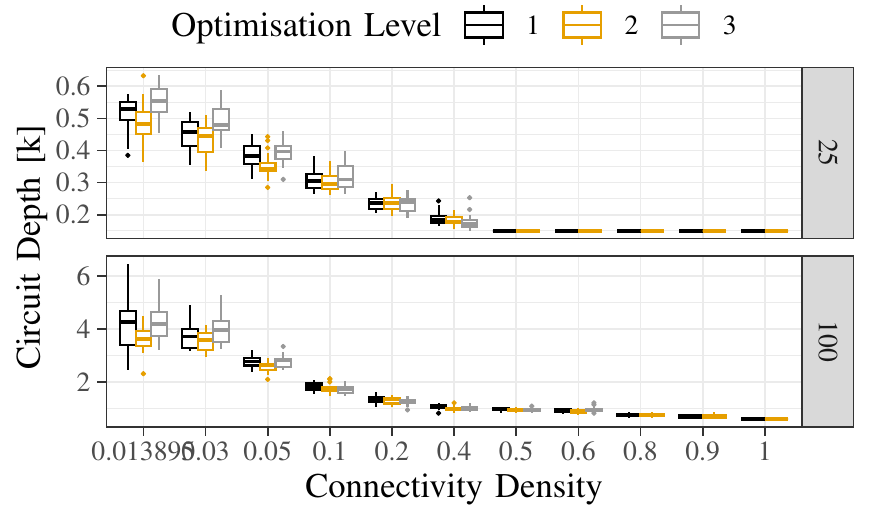}\vspace*{-1em}
  \caption{Distribution of circuit depth for two instance sizes
    (25 and 100 qubits) over \cd for varying optimisation levels obtained
    with the Qiskit compiler 0.41.1,
    see the replication package on the \suppweb for details) for the
    travelling salesperson problem.}%
  \label{fig:optimisation}
\end{figure}

Since the circuit mapping (transpilation) problem is known to be
\CC{NP}-complete by itself (see, \eg,~\cite{Paler:2021}), it is
unavoidable to use approximation techniques that cannot guarantee
optimal results in feasible time, and therefore require precise
characterisation. In particular, there is the risk that the technical
choice of optimisation level could impact our general conclusions; likewise,
different compiler/mapping approaches could influence behaviour.

Consider Figure~\ref{fig:optimisation}, which compares mapping results
with different optimisation levels for medium and large problem instances
requiring 25 and 100 qubits for the TSP (identical observations can be
made for the other subject problems). All levels follow the exponential
decrease pattern, with relatively small improvements of optimisation
level 3, although it is also clear that the highest optimisation level
does not guarantee smallest circuits, neither averaged nor overall.
As the highest optimisation level implies considerably increased simulation
times (days instead of hours), we find our choice of optimisation
level 2 justified. While there are many other approaches to circuit
compilation that we cannot compare in detail in the scope of this work, the results of Salm~\etal~\cite{Salm:2021}, together with results that
take differences for mapping practical problems between the most widespread
compilers into account~\cite{schoenberger:23:leap}, indicate that the
risk of observing a qualitatively different scaling behaviour is absolute
minor, though.

\subsubsection{Influence of Instance Properties}
\begin{figure}[htbp]
    \includegraphics{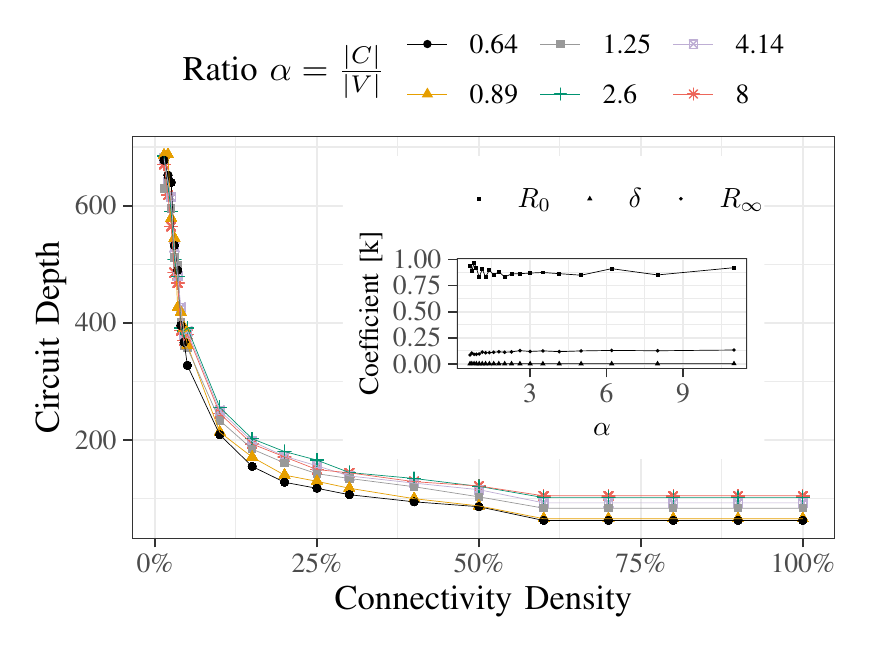}\vspace*{-1.5em}
    \caption{Empirically observed circuit depth degradation
    for MAX-3SAT on 36 variables, with varying
    degrees of \(\alpha\) (outer plot; see the main text
    for an explanation of this parameter), and the coefficients
    obtained for Eq.~\ref{eq:nls_model} (inset). The outer plot shows mean circuit depths
    obtained with 20 samples per data point and omit ranges
    to reduce visualisation clutter, while the
    model fit is obtained from the full data set.
    Connecting lines are used to guide the 
    eye.}\label{fig:sat_instances_depth} 
\end{figure}

Quantum circuits for a given problem are constructed using a uniform mechanism
that depends on problem size, but also on the properties of the instance itself.
As the observed exponential decrease in circuit depth might depend on the
latter properties, we explore a varying set of parameters for problem
MAX-3SAT. Boolean satisfiability is known to exhibit marked differences
in computational complexity depending on the ratio \(\alpha=|C|/|V|\) 
between the number of variables \(|V|\) and clauses \(|C|\)
(see, \eg, Refs~\cite{krueger:20:icse}). Fig.~\ref{fig:sat_instances_depth} 
shows how the circuit depth decreases with increasing \cd for random instances
of the problem that are constrained to a given value of \(\alpha\).
We scan across values of \(\alpha\) that represent
regions where instances are either trivially to solve by guessing
(\(\alpha \in [0,3.5]\)) or finding contradictory assignments that
show non-solubility (\(\alpha \in [4.9,11]\)),
as well as the region around \(\alpha\approx 4.2\) that is known to
contain computationally hard problem instances.

The inset shows that all three model coefficients \(R_{0}, R_{\infty}\) and
\(\delta\) as given in Eq.~\ref{eq:nls_model} are are in good agreement with
a constant value for each across the whole spectrum of \(\alpha\),
indicating no influence of the specific instance. Consequently,
we deem this risk minor to negligible. For this particular assessment,
it was not necessary to test other problem sizes, as 
\(\alpha\) determines the complexity of the Max3Sat problem.

\subsubsection{Influence of Backend Size}
\begin{figure}[htbp]
  \includegraphics{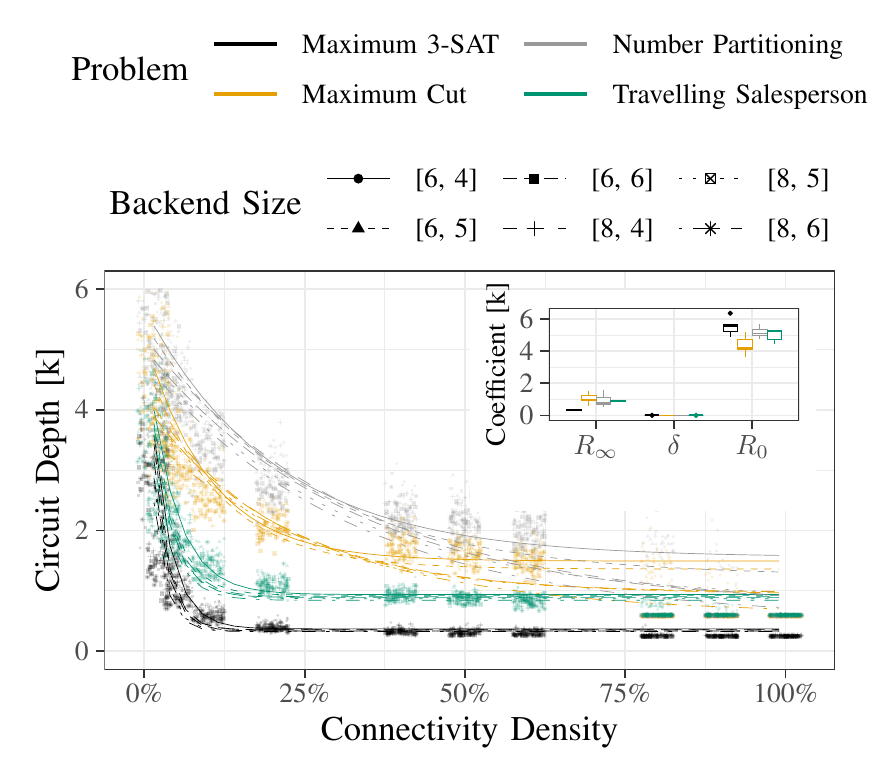}\vspace*{-1em}
  \caption{Empirical observations for circuit depth for 
  problems with constant size of 36 qubits with varying backend
  sizes (geometrically extrapolated from the
  IBM-Q heavy-hex backend), together with nonlinear
  regression fits of the model in Eq.~\ref{eq:nls_model}
  (we augment the empirical observations with a slight
  horizontal jitter to reduce overplotting. Inset: Distribution
  of regression coefficients.}
  \label{fig:backend_analysis}
\end{figure}

Our numerical experiments are performed using a constant
backend size of \(127\) qubits, so the ratio of the problem
size to the backend size varies. If the problem is much smaller
than the backend, several mappings of logical to physical qubits
are possible. We empirically study whether this influences
the circuit depth by uniformly scaling the backend size, 
while retaining the connectivity structure of the heavy-hex
geometry, which is composed of a certain number of rows
and columns containing interconnected rings.
Figure~\ref{fig:backend_analysis} shows circuit depths for varying backend sizes, whose
geometry has been consistently extended from the
IBM-Q Washington architecture with 127 qubits. Backend sizes
are specified in the form \([n,m]\), where \(n\) denotes
the number of rows, and \(m\) the number of columns
(details in the replication package on the \suppweb). 
This creates backends ranging from 143 (\(6\times 4\))
to 297 (\(8\times 6\)) qubits.

For none of the problems, the exponential decrease for increasing
\cd changes; there is practically no influence of the increased 
backend size for Max 3-SAT and TSP. For MaxCut and NumPart, the
asymptotes vary moderately depending on backend size, yet the 
differences are only relevant for \cds exceeding \(50\%\). 
However, since the models \emph{overestimate} the \cds compared to
empirical observations, we err---if at all---on the side of caution. 
These observations are also backed by the regression model coefficients,
whose distribution for each subject problem is shown in the inset.
Especially the rate parameter is extremely narrowly distributed,
which means that the exponential decrease with increasing \cd is
independent of the base backend size. Consequently, we deem this threat minor.

\section{Discussion \& Outlook}\label{sec:discussion}
In conclusion, our results show that an all-to-all connectivity
is not necessary to achieve near-optimal circuit depth for all subject problems.
Even small changes to the density can lead to significant improvements, particularly
across different problem sizes and types. 
Lower connectivity between qubits, lower circuit depth and lower gate counts 
can help scale quantum systems, as the number of interactions between
qubits, the number of quantum gates required to execute algorithms and the overall complexity of the system is decreased. 
Therefore it becomes easier
to maintain coherence and reduce the probability of errors and decoherence, 
which are crucial factors in building scalable quantum computing systems~\cite{inproceedings_reilly}.
We have identified an underlying effective model, which exhibits
an exponential decrease in circuit depth
with increasing connectivity uniformly across all instance sizes.
This suggest that our findings may be applicable to other problem
domains. Our results also point towards the construction of better
problem-adapted QPUs as a possible step towards practical applications
of quantum computing. The fact that all problems demonstrate a consistent exponential decrease in circuit depth as connectivity density rises is a highly encouraging and promising observation. This trend is true for all investigated problems.
Further research is required to explore the full
potential of our findings and understand the optimal topologies as well
as the effects on scalability for specific problems and problem classes. 
This includes a comprehensive analysis of the effects of 
noise and different topology layouts. Moreover, it is important to
incorporate more refined physical models that better capture the 
physical trade-offs involved. Finally, it is important to emphasise
the importance of hardware-software co-design for achieving 
scalability in quantum computing. As we continue to explore new 
algorithms and applications, it will be necessary to develop hardware
and software in tandem to ensure that they are optimised for each other. 

\noindent\textbf{Acknowledgements} This work is supported by the German 
Federal Ministry of Education and Research within the funding program \emph{Quantentechnologien -- von den Grundlagen zum Markt}, contract number 13N16092.

\FloatBarrier\clearpage
\nocite{*}
\bibliographystyle{IEEEtran}
\bibliography{main}

\end{document}